\documentclass[12pt]{article}
\usepackage{bbm}
\usepackage{color}
\usepackage{authblk}
\usepackage{latexsym}
\usepackage{epsfig,amssymb,euscript, mathrsfs,cite}
\usepackage{amsmath}
\usepackage{slashed}
\usepackage{mathrsfs} 
\usepackage{enumerate}
\definecolor{MyDarkBlue}{rgb}{0.15,0.15,0.45}
\usepackage[linktocpage=true]{hyperref}
\hypersetup{
colorlinks=true,
citecolor=MyDarkBlue,
linkcolor=MyDarkBlue,
urlcolor=MyDarkBlue,
pdfauthor={},
pdftitle={},
pdfsubject={hep-th}
}
\newsavebox{\ns}
\newsavebox{\dbrane}
\newsavebox{\dbshort}

\def\be{\begin{equation}}
\def\ee{\end{equation}}
\def\bea{\begin{eqnarray}}
\def\eea{\end{eqnarray}}

\newcommand{\nn}{\nonumber\\}

\newcommand\R{\mathbb{R}}
\newcommand\Z{\mathbb{Z}}

\newcommand\C{\mathbb{C}}

\newcommand\diff{\mathrm{d}}

\newcommand{\ex}{\mathrm{e}}

\newlength{\sswidth}

\numberwithin{equation}{section}       % equation numbers in each section

\newcommand{\qprime}{k}

\newcommand{\pp}{P}
\newcommand{\qq}{Q}

\begin{document}

\begin{titlepage}

\begin{flushright}
Imperial/TP/2021/JG/03\\
\end{flushright}

\vskip 1cm

\begin{center}

%\today

{\Large \bf M5-branes wrapped on a spindle}

\vskip 1cm

\vskip 1cm
{Pietro Ferrero$^{\mathrm{a}}$, Jerome P. Gauntlett$^{\mathrm{b}}$,  Dario Martelli$^{\mathrm{c,d,e}}$, and James Sparks$^{\mathrm{a}}$}

\vskip 0.5cm

${}^{\,\mathrm{a}}$\textit{Mathematical Institute, University of Oxford,\\
Andrew Wiles Building, Radcliffe Observatory Quarter,\\
Woodstock Road, Oxford, OX2 6GG, U.K.\\}

\vskip 0.2 cm

${}^{\mathrm{b}}$\textit{Blackett Laboratory, Imperial College, \\
Prince Consort Rd., London, SW7 2AZ, U.K.\\}

\vskip 0.2cm

${}^{\mathrm{c}}$\textit{Dipartimento di Matematica ``Giuseppe Peano'', Universit\`a di Torino,\\
Via Carlo Alberto 10, 10123 Torino, Italy}

\vskip 0.2cm

${}^{\mathrm{d}}$\textit{INFN, Sezione di Torino \&}   ${}^{\mathrm{e}}$\textit{Arnold--Regge Center,\\
 Via Pietro Giuria 1, 10125 Torino, Italy}

\end{center}

\vskip 0.5 cm

\begin{abstract}
\noindent  
 We construct supersymmetric $AdS_5\times \Sigma$ solutions 
of $D=7$ gauged supergravity, where $\Sigma$ is a two-dimensional orbifold known as a spindle. 
These uplift on $S^4$ to solutions of 
$D=11$ supergravity which have orbifold singularites. 
We argue that the solutions are dual to  $d=4$, $\mathcal{N}=1$ SCFTs that 
arise from $N$ M5-branes wrapped on a spindle, 
embedded as a holomorphic curve inside a Calabi-Yau three-fold. 
In contrast to the usual topological twist solutions, 
the superconformal R-symmetry mixes with the isometry of the spindle in the IR, 
and we verify this via a field theory calculation, as well as reproducing 
the gravity formula for the central charge. 

\end{abstract}

\end{titlepage}

\pagestyle{plain}
\setcounter{page}{1}
\newcounter{bean}
\baselineskip18pt

\tableofcontents

\newpage

\section{Introduction}\label{sec:intro}

A rich landscape of SCFTs arise from the low-energy limit of branes wrapping supersymmetric cycles.
Furthermore, important features of these SCFTs can be elucidated holographically by constructing
suitable supergravity solutions. A well-studied framework, starting with \cite{Maldacena:2000mw,Gauntlett:2000ng,Gauntlett:2001qs}, is to construct
supergravity solutions associated with M5-branes, M2-branes or D3-branes wrapping supersymmetric 
cycles with supersymmetry realized via a (partial) topological twist. The topological twist couples the
field theory to external R-symmetry currents which effectively (partially) cancels the spin connection on the world volume 
of the wrapped brane, thus allowing supersymmetry to be realized \cite{Witten:1988ze}.
In particular, the standard topological twist is associated with constant Killing spinors on the cycle upon which the brane is wrapped. 
Geometrically, thinking of the branes
as wrapping calibrated cycles inside manifolds with special holonomy, the topological twist
can also be viewed as arising from the structure of the normal bundle \cite{Bershadsky:1995qy}, which then allows for such constant sections.

A powerful way to construct such supergravity solutions is to construct $AdS\times \Sigma^{(d)}$ solutions of a gauged supergravity theory, where $\Sigma^{(d)}$ is the $d$-dimensional cycle upon which the branes are wrapped, and then uplift to $D=10$ or $11$. 
 In the standard constructions, $\Sigma^{(d)}$ is tightly constrained: for example in the two-dimensional case\footnote{The constructions of \cite{Gauntlett:2000ng} include examples of
higher-dimensional cycles where the metric does not have constant curvature.}, 
the case of most interest in this paper, $\Sigma^{(2)}$ is a Riemann surface of genus $\mathtt{g}$ with a constant curvature metric.
In recent work, it has been shown that there are fundamentally new constructions where this is not the case, and $\Sigma^{(2)}$ is instead a ``spindle" \cite{Ferrero:2020laf}.

A spindle has the same topology as a two-sphere, but there are orbifold singularities at the north and south poles which are associated with quantized conical deficit angles. More precisely, we define a spindle as the weighted-projective space 
$\Sigma \equiv \mathbb{WCP}^1_{[n_-,n_+]}$, specified by two relatively prime integers $n_\pm\in\mathbb{N}$. In \cite{Ferrero:2020laf} solutions of $D=5$ minimal gauged supergravity of the form $AdS_3\times \Sigma$, dual to $\mathcal{N}=(0,2)$ SCFTs in $d=2$, were constructed.
After carefully uplifting on five-dimensional Sasaki-Einstein manifolds in the regular class, one obtains solutions of type IIB supergravity of the form $AdS_3\times  \Sigma\times SE_5$ which, surprisingly, are free from all conical singularities
and, furthermore, were constructed earlier in \cite{Gauntlett:2006ns}. The gravity solution suggests that the $d=2$ dual SCFT should
be viewed as arising from the $d=4$ SCFT dual to the $AdS_5\times SE_5$ solution, wrapping it on the spindle and then flowing to the IR. Further support for this interpretation was provided by showing that the central charge of the $d=2$ SCFT calculated from the gravity solution exactly matches the result obtained from the anomaly polynomial for the $d=4$ SCFT, after reducing on the spindle and then employing the $c$-extremization procedure of \cite{Benini:2012cz}. An interesting feature is that
the R-symmetry of the $d=2$ SCFT arises from a mixing of the R-symmetry of the $d=4$ SCFT with the rotational symmetry of the spindle.

The results of \cite{Ferrero:2020laf} immediately suggest that there could be a rich landscape of new supergravity solutions associated with branes wrapping spindles and higher-dimensional orbifolds. In fact further interesting examples have already appeared.
In \cite{Hosseini:2021fge,Boido:2021szx} the solutions of \cite{Ferrero:2020laf} were generalized from minimal $D=5$ gauged supergavity to the STU model, which has three Abelian gauge fields. The solutions can be uplifted on the five-sphere to obtain solutions of type IIB supergravity describing D3-branes wrapping the spindle, which are again completely regular, and were first constructed in \cite{Gauntlett:2006ns}. 
Furthermore the central charge of the $d=2$ SCFT obtained from the gravity solution agrees exactly 
with the associated field theory calculation using the anomaly polynomial of $\mathcal{N}=4$ SYM theory in $d=4$.
In the solutions describing D3-branes wrapped on spindles in both \cite{Ferrero:2020laf} and \cite{Hosseini:2021fge,Boido:2021szx}, supersymmetry is not being realized by the topological twist; for example, it was shown by explicit construction in \cite{Ferrero:2020laf} that 
the Killing spinors are not constant on $\Sigma$, and moreover are in fact sections of non-trivial bundles over $\Sigma$.

In \cite{Boido:2021szx} a sub-class of the $D=5$ solutions of the STU model were uplifted to $D=11$ supergravity on $\Sigma_\mathtt{g}\times S^4$, where $\Sigma_\mathtt{g}$ is a Riemann surface of genus $\mathtt{g}>1$.
These are regular $D=11$ solutions, first found in \cite{Gauntlett:2006qw}, that describe M5-branes wrapped on a product four-cycle 
$\Sigma\times \Sigma_\mathtt{g}$. In these solutions 
supersymmetry is being realized by a standard topological twist on $\Sigma_\mathtt{g}$, but not so for the spindle $\Sigma$.

Another generalisation, describing M2-branes wrapping spindles, was discussed in \cite{Ferrero:2020twa}. A class of $AdS_2\times \Sigma$ solutions were constructed in minimal $D=4$ gauged supergravity. After uplifting these solutions to $D=11$ using $SE_7$ in the regular class, one again obtains regular solutions, first found in \cite{Gauntlett:2006ns}, and once again supersymmetry is not being realized by a topological twist. 
This class of solutions also exhibits several notable features. 
Firstly, the solutions can be generalized to include a rotation parameter.
Secondly, the solutions were shown to arise as the near-horizon limit of a class of $D=4$ accelerating black holes, with the conical defects of the spindle being directly associated with the acceleration. 
After uplifting to $D=11$, the black hole solutions can be viewed as describing a flow across dimensions from the $d=3$ SCFTs, dual to the $AdS_4\times SE_7$ solutions, down to the quantum mechanics dual to the $AdS_2$ solutions. A curious feature is that in these specific UV completions
the special case of vanishing rotation is associated with a standard topological twist, but in a strange limiting way where the conformal boundary splits into two halves with a different topological twist on each half. A detailed field theory comparison has yet to be made for these wrapped M2-brane solutions.

In this paper we continue to explore the new landscape, by presenting a new class of solutions describing M5-branes wrapping spindles. Once again, while we find some similarities with the examples discussed above, we also find some new features. In contrast to D3-branes and M2-branes, we do not find any solutions describing M5-branes wrapping spindles in minimal $D=7$ gauged supergravity. Here, instead, we construct new $AdS_5\times \Sigma$ solutions of
$D=7$ gauged SUGRA coupled to two $U(1)$ gauge fields, which are dual to $\mathcal{N}=1$ SCFTs in $d=4$. The local solutions were in fact first obtained by a simple double analytic continuation \cite{Lu:2003iv} of the supersymmetric, static R-charged black holes
studied in \cite{Cvetic:1999ne,Cvetic:1999xp, Liu:1999ai}. 

Interestingly, we find that supersymmetry is being realized in a new way. On the one hand, the spin connection of the spindle is not equal to the R-symmetry gauge fields, and the associated Killing spinor is not constant on the spindle.
This is similar to the case of the wrapped  D3-branes and M2-branes. However, on the other hand, in contrast to those cases we find that the 
(integrated) R-symmetry flux is equal to the Euler character of the spindle. In the usual topological twist 
this follows as a corollary of the local identification of the spin connection with the R-symmetry gauge fields. Thus,
the way in which supersymmetry is being realized for the $AdS_5\times \Sigma$ solutions might be referred to as
a ``global topological twist," or perhaps described as ``topologically a  topological twist''!

After uplifting on a four-sphere, we obtain solutions of $D=11$ supergravity but now they are still singular.
It is not yet clear how these orbifold singularities should be interpreted and/or resolved but we provide evidence
that the $D=11$ solutions are holographically describing M5-branes wrapping spindles. 
In particular, we show that the central charge of the $d=4$ SCFT as calculated from the gravity side
agrees exactly with the associated field theory computation using 
the approach of \cite{Ferrero:2020laf}, now involving
the anomaly polynomial of the $\mathcal{N}=(0,2)$ SCFT 
in $d=6$, and the $a$-extremization principle of \cite{Intriligator:2003jj}.
We will argue that the solutions should be viewed as arising from the IR limit
of M5-branes wrapping a spindle, holomorphically embedded inside a Calabi-Yau three-fold.

Our construction of the new wrapped M5-brane solutions is complementary to that in \cite{Bah:2012dg}, which considered M5-branes wrapped on a Riemann surface $\Sigma_\mathtt{g}$, equipped with a constant curvature
metric and with the standard topological twist. In both cases, the global description of the geometry is given in terms of M5-branes wrapped on the zero section of the total space of vector bundles  $\mathcal{O}(-p_1)\oplus  \mathcal{O}(-p_2)\to
 \mathcal{M}_\chi$, with $(p_1+p_2)/n_-n_+=\chi$ so that the total first Chern class vanishes. In  \cite{Bah:2012dg}  $\mathcal{M}_\chi=\Sigma_\mathtt{g}$ is a Riemann surface with Euler number $\chi=2(1-\mathtt{g})$,
while  here  $\mathcal{M}_\chi=\Sigma$ is the spindle.\footnote{Formally, setting $n_+=n_-=1$ we recover the case of $\Sigma=\Sigma_0=S^2$. However,
this supergravity solution does not fall in our class, as explained earlier, and was in fact found in  \cite{Cucu:2003bm}.} A
crucial difference is that while genus $\mathtt{g}$ Riemann surfaces admit constant curvature metrics, spindles do not admit such metrics and this is associated with the distinct realization of supersymmetry. If we set $n_-=n_+=1$ then our 
construction of the $D=11$ supergravity solutions degenerates. Nevertheless, we find that formally setting 
$n_-=n_+=1$ in our final expression for the central charge, given in \eqref{ftheorycentchge}, then we precisely recover
the expression for the central charge of \cite{Bah:2012dg} for the case of the standard topological twist and genus $g=0$. 
Something similar happens for accelerating black hole solutions associated with M2-branes wrapping ``spinning spindles" \cite{Ferrero:2020twa}.

The outline of the paper is as follows. In section \ref{sec:soln} we construct 
the $AdS_5\times \Sigma$ solutions, in particular analysing and solving 
the conditions required to have a smooth orbifold metric on $\Sigma=\mathbb{WCP}^1_{[n_-,n_+]}$, 
with properly quantized magnetic fluxes. In section~\ref{sec:uplift} we uplift 
these solutions to M-theory, making contact with M5-branes 
wrapping $\Sigma$ in a Calabi-Yau three-fold, and computing the $a$ central charge 
in gravity. Section \ref{sec:anom} reduces the anomaly polynomial 
of the M5-branes to $d=4$, and we compute the exact superconformal R-symmetry 
and $a$ central charge in field theory using $a$-maximization, finding 
precise agreement with the gravity result. 
Section \ref{discuss:sec} concludes with a discussion. The appendix 
includes details of certain integrals that appear in the main text.

\section{$AdS_5\times \Sigma$ solutions}
\label{sec:soln}

In this section we construct a family of supersymmetric $AdS_5\times \Sigma$ solutions 
of $D=7$ gauged supergravity, where $\Sigma=\mathbb{WCP}^1_{[n_-,n_+]}$ is a 
spindle, parametrized by arbitrary coprime positive integers $n_->n_+$. 
The gauged supergravity theory has two Abelian gauge fields $A_i$, $i=1,2$, and their 
magnetic fluxes through $\Sigma$ are characterized by integers $p_i$, 
which satisfy the constraints $p_1+p_2=n_-+n_+$ with $p_1\times p_2<0$. 

\subsection{Local form of the solutions}\label{sec:local}

We are interested in solutions of a truncation of $D=7$ gauged supergravity theory that keeps two $U(1)$ gauge fields, $A_i$, and two real scalar fields, $\vec{\varphi}=(\varphi_1,\varphi_2)$. The Lagrangian is given by
\begin{align}
\mathcal{L}\, = \, R-g^2\mathcal{V}-\frac{1}{2} \partial_{\mu}\vec{\varphi}\cdot \partial^{\mu}\vec{\varphi}   -\frac{1}{4}\sum_{i=1}^2X_i^{-2}\,F_{i\; \mu\nu} F_i^{\mu\nu}\,,
\end{align}
where $F_i=\diff A_i$,  $X_i=\ex^{-\frac{1}{2}\vec{a}_i\cdot \vec \varphi}$, $i=1,2,$ and the vectors $\vec{a}_i$ are given by
\begin{align}
\vec{a}_1\, = \, \Big(\sqrt{2},\, \sqrt{\tfrac{2}{5}}\Big)\,, \qquad
\vec{a}_2\, = \, \Big(-\sqrt{2},\, \sqrt{\tfrac{2}{5}}\Big)\,.
\end{align}
The scalar potential is 
\begin{align}
\mathcal{V}\, = \, \tfrac{1}{2} X_1^{-4}\,X_2^{-4}-2X_1^{-1}\,X_2^{-2}-2X_1^{-2}\,X_2^{-1}-4 X_1\,X_2\,,
\end{align}
and in the rest of this paper  we will set  $g=1$. While not itself a consistent truncation of $D=11$ supergravity, it was shown in \cite{Cvetic:1999xp} that solutions with
$F_1\wedge F_2=0$, of relevance in this paper, can be uplifted on an $S^4$ to obtain solutions
of $D=11$ supergravity.

The supersymmetric $AdS_5\times \Sigma$ solutions of interest are given by 
\begin{align}\label{7dmetric}
\diff s^2_7&  \, = \, (y \pp(y))^{1/5}\left[ \diff s^2_{AdS_5}+   \frac{y}{4\qq(y)}\diff y^2+ \frac{\qq (y)}{\pp(y)}\diff z^2\right]\, , \nonumber\\
 A_i & \, = \,  \frac{q_i}{h_i(y)}\diff z \, ,  \qquad X_i (y)  \, = \,   \frac{\left(y\pp(y) \right)^{2/5} }{h_i(y)}\, ,
\end{align}
where  $\diff s^2_{AdS_5}$ is the unit radius metric on $AdS_5$, and 
\begin{align}\label{metricfunctions}
h_i (y) & \, = \,  y^2 + q_i\, , \nonumber\\
\pp(y) & \, = \,  h_1 (y) h_2(y) \,  = \,   (y^2+ q_1) (y^2+q_2) \, ,  \nonumber\\
\qq(y) & \, = \,  -y^3+ \tfrac{1}{4}    \pp(y) \, = \,   -y^3+ \tfrac{1}{4}   (y^2+ q_1) (y^2+q_2)\, ,
\end{align}
with $q_1,q_2$ two real parameters. These solutions can be simply obtained by doing an analytic  continuation \cite{Lu:2003iv} of the supersymmetric, static R-charged black holes constructed in \cite{Cvetic:1999ne,Cvetic:1999xp, Liu:1999ai}. 

\subsection{Global analysis and magnetic fluxes}\label{sec:global}

We are interested in determining the conditions on the parameters $q_i$ so that the two-dimensional metric 
\begin{align}
\label{myspindle}
\diff s^2_\Sigma & \, \equiv \,  \frac{y}{4\qq(y)}\diff y^2+ \frac{\qq (y)}{\pp(y)}\diff z^2\,,
\end{align}
is a smooth metric on a spindle $\Sigma=\mathbb{WCP}^1_{[n_-,n_+]}$, with $n_\pm$ coprime positive integers with $n_->n_+$.
Necessary conditions for this to be the case are  that 
$\qq(y)>0$, $h_1(y)>0$, $h_2(y)>0$ in an interval $y\in (y_a,y_b)$, where $y_a,y_b$ are two consecutive real roots of $\qq(y)=0$, with $y_b>y_a>0$.  Furthermore,  since the coefficient of $y^4$ in $\qq(y)=0$ is positive,  the presence of a double root would necessarily imply that \eqref{myspindle} is not a metric on a compact space.  All in all these conditions imply that we need four real roots for $\qq(y)=0$,  with at least three of them positive.  To show that this is indeed possible,  one can simply study the signs of the roots of $\qq(y)$ by Descartes's rule of signs.  We consider various cases separately:
\begin{itemize}
\item when $q_1>0$,  $q_2>0$ there are no negative roots and either 0 or 2 positive roots.  So the number of real roots is at most 2,  which is not allowed.
\item when $q_1<0$,  $q_2<0$ there are either 0 or 2 negative roots,  and either 0 or 2 positive roots.  Hence,  when they are all real the two middle roots have opposite sign,  which again is not allowed.
\item when $q_1\,q_2<0$ there is always 1 negative root,  while the positive roots can be either 1 or 3.  Hence,  in this case we can have four real roots,  with the middle two being both positive: this is when the metric is regular.
\item when $q_1\,q_2=0$ there is a double root at $y=0$, which would not give rise to a complete metric on a compact space.\footnote{This case is expected to preserve twice as much supersymmetry.}
\end{itemize}
Hence, continuing with
\begin{align}
q_1\,q_2<0\,,
\end{align}
it is possible to take $y\in [y_a,y_b]$, with $\qq(y)>0$ for $y\in (y_a,y_b)$, and $y_b>y_a>0$.  Without loss of generality one can take $q_1>0$ and $q_2<0$,  which clearly implies that $h_1(y)>0$.  Then  we note that we can write
\begin{align}
\qq(y)\, = \, -y^3+\tfrac{1}{4}\,h_1(y)\,h_2(y)\,,
\end{align}
and since $\qq(y)>0$ for $y\in (y_a,y_b)$ we have 
\begin{align}
h_2(y)\, > \, \frac{4\,y^3}{h_1(y)}\,,
\end{align}
which is clearly positive in the given range.  This also shows that $\pp(y)>0$ as $\pp(y)=h_1(y)\,h_2(y)$.  We conclude that a regular solution is possible only when $q_1\,q_2<0$,  and all four roots are real. We then take $a=2$, $b=3$, so that $y\in [y_2,y_3]$.

As $y$ approaches the two end-points of the interval, $y\to y_2,y_3$, the metric $\diff s^2_\Sigma$ in \eqref{myspindle} takes the approximate form
\begin{align}
\diff s^2_\Sigma& \, \simeq \,    \diff \rho^2  +   \rho^2  \frac{\qprime_i^2}{4y_i^4} \diff z^2 \, , 
\end{align}
where
\begin{align}
  \qprime_i \, \equiv \, \qq'(y_i)\,  =  \,  \frac{(q_1+q_2)y_i }{2}+ (y_i-3) y_i^2\, ,
  \end{align}
  and we have $\qprime_2>0$, $\qprime_3<0$. Demanding that $z$ is a periodic coordinate with period $\Delta z$ given by
    \begin{align}
    \label{systemreg}
 \frac{k_2\Delta z }{2y_2^2} \, = \,  \frac{2\pi}{n_+}\, , \qquad \qquad  \frac{\qprime_3 \Delta z}{2y_3^2}  \, = \,  - \frac{2\pi}{n_-} \,  ,
\end{align}
with $n_\pm$ coprime positive integers, then ensures that we have a metric on a spindle which is regular 
everywhere, apart from the conical deficit angles $2\pi(1-\frac{1}{n_\pm})$ at the poles $y=y_2$, $y_3$, which 
are orbifold singularities.

Using the following expression for the Ricci scalar of the metric (\ref{myspindle})
%\begin{align}
%\sqrt{g_\Sigma}\,R_\Sigma \, = \, 2\frac{\mathrm{d}}{\mathrm{d}y} \left(\frac{\qq(y)\pp'(y)-\qq'(y)\pp(y)}{y^{1/2}\,\pp(y)^{3/2}}\right)\, ,
%\end{align}
\begin{align}
\sqrt{g_\Sigma}\,R_\Sigma \, = \, 2 \left(\frac{\qq\pp'-\qq'\pp}{y^{1/2}\,\pp^{3/2}}\right)'\, ,
\end{align}
 one can immediately check that the Euler number
\begin{align}\label{Euler}
\chi(\Sigma) \, = \,  \frac{1}{4\pi}\int_{\Sigma} R_\Sigma \mathrm{vol}_\Sigma  \, =  \, \frac{n_-+n_+}{n_-n_+}\, ,
\end{align}
takes the correct value for the spindle. 

Having analysed the global conditions required for the metric of the supersymmetric $AdS_5\times \Sigma$ solutions, 
next we turn to the 
appropriate quantization conditions for the magnetic fluxes threading the spindle $\Sigma$. 
Specifically, in order that $A_i$ are well-defined connection one-forms on $U(1)$ bundles over the spindle
we require that\footnote{Note that the gauge fields are normalized so that the $D=7$ Killing spinors have charge 1/2 with respect to both
$A_i$, as one can check from e.g. \cite{Lu:2003iv}.}
\begin{align}
\label{fluxcond}
P_i & \, \equiv \,  \frac{1}{2\pi}\int_\Sigma \diff A_i \, = \,  \frac{p_i}{n_- n_+}\,,\qquad p_i\in\mathbb{Z}\,,
\end{align}
(e.g. see appendix A of \cite{Ferrero:2020twa}). 
The integrals can be performed straighforwardly and we find that we must demand 
\begin{align}
P_i & \, = \,  \frac{\Delta z}{2\pi}q_i\left(\frac{1}{h_i(y_3)} - \frac{1}{h_i(y_2)}\right) \,  = \,   \frac{p_i}{n_-n_+}\, .
\label{systemflux}
\end{align}
It is next 
illuminating to calculate the 
total flux for the R-symmetry of the dual $d=6$, $\mathcal{N}=(0,2)$ SCFT, which is given by $P_1+P_2$, as we will explain later. 
 A simple computation gives   
\begin{align}
P_1+P_2  &  \,  = \,  \frac{\Delta z}{2\pi}\left[\frac{(q_1+q_2)y_3^2+2q_1q_2}{4y_3^3}  -    \frac{(q_1+q_2)y_2^2+2q_1q_2}{4y_2^3}   \right]\, , 
\end{align}
where  we used $h_1(y_i)h_2(y_i)=4y_i^3$.
After using the  identity 
\begin{align}
\label{usefulid}
\frac{\qprime_i}{2y_i^2} \, = \,  \frac{1}{2} - \frac{(q_1+q_2)y_i^2+2q_1q_2}{4y_i^3}\, , 
\end{align}
which can be proved using $\qq(y_i)=0$, we find the remarkable result that
\begin{align}\label{sumcharges}
P_1+P_2  
 & \, =  \, \frac{\Delta z}{2\pi}\left( -\frac{\qprime_3}{2y_3^2}+\frac{\qprime_2}{2y_2^2}     \right) \, = \,  \frac{n_-+n_+}{n_-n_+} \, = \, \chi(\Sigma)\, . 
\end{align}
This result can be contrasted with analogous solutions describing D3-branes and M2-branes wrapping spindles
in \cite{Ferrero:2020laf} and \cite{Ferrero:2020twa}, respectively, where the total R-symmetry flux though the spindle was
instead given\footnote{This might be referred to as an ``anti-topological twist'', due to the relative minus 
sign compared with the Euler number in \eqref{sumcharges}. 
We also note that in \cite{Ferrero:2020laf},\cite{Ferrero:2020twa} the R-symmetry gauge field was normalized such that the $D=5,4$ gauged supergravity spinors carried unit charge, whereas here, with $A^R\equiv A_1+A_2$, it is normalized to have charge $1/2$. 
%This accounts for the
%1/2 appearing on the right hand side of eq (7) of \cite{Ferrero:2020laf} and the factor in (4.31) of \cite{Ferrero:2020twa}.
} by $(n_--n_+)/(n_- n_+)$. In fact \eqref{sumcharges} 
might naively be
identified with a ``topological twist'' since it is a corollary of the usual topological twist when there is a local
identification of the spin connection with the R-symmetry gauge fields. However, there 
are a number of differences between our 
construction and the more standard construction of 
$AdS \times \Sigma_\mathtt{g}$ solutions of gauged supergravity, where $\Sigma_\mathtt{g}$ is a Riemann surface with genus $\mathtt{g}$ \cite{Maldacena:2000mw}.
In standard constructions the topological twist condition leads to a constant curvature metric on  
 $\Sigma_\mathtt{g}$ and Killing spinors that are constant on $\Sigma_\mathtt{g}$. 
 Instead in our $AdS_5\times \Sigma$ solutions
the metric on $\Sigma$ does not have constant curvature, and indeed this is necessary 
as $\Sigma=\mathbb{WCP}^1_{[n_-,n_+]}$ does not admit a metric of constant curvature unless $n_-=n_+=1$, when 
$\Sigma=S^2$. Furthermore, the Killing spinors, which can be obtained by analytic continuation from those given in
\cite{Liu:1999ai}, are not constant on $\Sigma$. The global condition \eqref{sumcharges} might be referred to
as ``topologically a topological twist".
It is clearly of interest to understand the reason for the different total R-symmetry twists on spindles
in the cases of D3-branes and M2-branes, versus the M5-brane construction in this paper. 
Finally, notice that \eqref{sumcharges} implies that the $p_i$ and $n_\pm$ satisfy the constraint
 \begin{align}\label{constraintpsen}
 p_2\, = \, n_-+n_+-p_1\,,
 \end{align}
which allows us to eliminate $p_2$, for example, in subsequent formulae.

\subsection{Solution of the regularity conditions}\label{sec:regular}

We now analyse the regularity conditions that are required to have good $AdS_5\times\Sigma$ solutions with properly quantized fluxes,  namely \eqref{systemreg} and \eqref{systemflux}. We would like to obtain expressions for the 
parameters $q_1$, $q_2$, as well as $\Delta z$, in terms of the spindle data $n_-,n_+$ and the integer $p_1$, recalling that $p_2$ can be obtained from the constraint \eqref{constraintpsen}.

The most straightforward approach is to solve the equation $\qq(y)=0$ for $y$ and replace the value of the roots in the relevant equations in the previous subsection. However, this is complicated in practice by the unwieldy explicit expressions for the roots of $\qq(y)$.  We hence follow a different strategy. 
We begin by writing
\begin{align}\label{qpolynomial}
\qq(y)\, = \, \frac{1}{4}(y-y_1)(y-y_2)(y-y_3)(y-y_4)\,,
\end{align}
where  by comparing with $\qq(y)$ given in \eqref{metricfunctions},  one can read off the constraints that the $y_i$ must satisfy.  We then have a set of seven equations for the unknowns $y_i$ ($i=1,...,4$),  $q_i$ ($i=1,2$) and $\Delta z$,  to be solved in terms of the three parameters $n_{\pm}$ and $p_1$.
Some of the unknowns can be trivially eliminated by solving linear equations,  that give
\begin{align}
y_4&\, = \, 4-y_1-y_2-y_3\,, \nonumber\\
q_2&\, = \, -q_1+4(y_1+y_2+y_3)-(y_1^2+y_2^2+y_3^2+y_1	y_2+y_1\,y_3+y_2	\,y_3)\,,\nonumber\\
%\Delta z&\, = \, \frac{16\pi\,y_2^2}{n_+\,(y_2-y_3)}\,\left[(y_2-y_1)(4-y_1-y_2-y_3)\right]^{-1}\,,\nn
\Delta z&\, = \, \frac{16\pi\,y_2^2}{n_+\,(y_2-y_3)(y_2-y_1)(-4+y_1+2y_2+y_3)}\,,
\end{align}
where the last equation is obtained from the first of \eqref{systemreg} after using \eqref{qpolynomial} to get $k_2$.
This leaves us with a system of four equations to be solved for $y_{1,2,3}$ and $q_1$,  namely
\begin{align}\label{equationsregularity}
0&\, =\, n_+  y_3^2\left(y_1-y_2\right) \left(y_1+2 y_2+y_3-4\right)-n_- y_2^2 \left(y_1-y_3\right) \left(y_1+y_2+2 y_3-4\right),\nonumber\\
%0&\, = \, p_1 \left(y_1-y_2\right) \left(y_1+2 y_2+y_3-4\right) \left(q_1 \left(y_2^2+y_3^2\right)+q_1^2+y_2^2 y_3^2\right)+8 n_- q_1 y_2^2 \left(y_2+y_3\right),\nonumber\\
0&\, = \, p_1 \left(y_1-y_2\right) \left(y_1+2 y_2+y_3-4\right)(q_1+y_2^2)(q_1+y_3^2)+8 n_- q_1 y_2^2 \left(y_2+y_3\right),\nonumber\\
0&\, = \, q_1^2+\left(y_1^2+y_2^2+(y_1+y_3)(y_2+y_3)-4(y_1+y_2+y_3)\right)\,q_1 \nonumber\\
& \quad\  -y_1 y_2 y_3 \left(y_1+y_2+y_3-4\right),\nonumber\\
%0&\, = \, -\left(y_2+y_3\right) y_1^2-\left(y_2^2+2 \left(y_3-2\right) y_2+\left(y_3-4\right) y_3\right) y_1-y_2 y_3 \left(y_2+y_3-4\right).\nn
0&\, = \, \left(y_2+y_3\right) y_1^2+\left(-4+y_2+y_3\right)\left(y_1y_2+y_1y_3+y_2 y_3\right) ,
\end{align}
where the first equation comes from the second of \eqref{systemreg} and the second equation comes from the first of
\eqref{systemflux}.
Remarkably,  it is possible to solve this system by solving only quadratic equations,  which allows to write the solution in a relatively compact form.   To do so,  one can solve the first equation for $y_1$ and the third for $q_1$.  The remaining two equations can be rearranged in such a way that one has a quadratic for $y_2-y_3$,  whose solution gives a quadratic equation for $y_2+y_3$.  After some massaging,  the result for $y_2,y_3$ reads
\begin{align}\label{y_{2,3}}
y_2&=\frac{3\,p_1\,p_2\,(5\,n_+-n_-+\mathtt{s})(\mathtt{s}+p_1+p_2)}{2\,(n_--p_1)(n_--p_2)[\mathtt{s}+2\,(p_1+p_2)]^2}\,, \nonumber \\
y_3&=\left. y_2\right|_{n_+\leftrightarrow n_-} \,,
\end{align}
while for $y_4,y_1$ we have
\begin{align}
y_{4,1}=&\frac{(5\,n_--n_++\mathtt{s})\,(5\,n_+-n_-+\mathtt{s})\,(p_1+p_2+\mathtt{s})}{24\,(n_--p_1)\,(p_2-n_-)\,[\mathtt{s}+2\,(p_1+p_2)]^2}\nonumber \\
&\qquad\times\left(\mathtt{s}+2\,(p_1+p_2)\pm\sqrt{(p_1+p_2+2\,\mathtt{s})^2-36\,n_-\,n_+}\right)\,,
\end{align}
where the upper sign in the last equation corresponds to $y_4$,  the lowest to $y_1$
and we have defined
\begin{align}\label{sdef}
\mathtt{s}\, &\equiv
\sqrt{7\,(p_1^2+p_2^2)+2\,p_1\,p_2-6\,(n_-^2+n_+^2)}\,,\nn
 &= \sqrt{(n_-+n_+)^2+12(n_--p_1)(n_+-p_1)}\,,
%=\sqrt{7\,(p_1^2+p_2^2)-6\,(n_-^2+n_+^2)+2\,p_1\,p_2}\,.
\end{align}
where we used the constraint \eqref{constraintpsen}: $p_2=n_-+n_+-p_1$.
We also obtain the following compact expressions for $q_1,q_2$,
\begin{align}\label{q_{1,2}}
q_1&=\frac{3\,p_1\,p_2^2\,(5n_--n_++\mathtt{s})\,(5n_+-n_-+\mathtt{s})\,(p_1-2\,p_2-\mathtt{s})\,(p_1+p_2+\mathtt{s})^2}{4\,(n_--p_1)^2\,(n_--p_2)^2\,[\mathtt{s}+2(p_1+p_2)]^4}\,, \nonumber \\
q_2&=\left. q_1\right|_{p_1 \leftrightarrow p_2}\,,
\end{align}
and for $\Delta z$:
\begin{align}\label{Deltaz}
\Delta z=\frac{[\mathtt{s}-(p_1+p_2)]\,[\mathtt{s}+2(p_1+p_2)]}{9\,n_-\,n_+\,(n_--n_+)}\,2\pi\,.
\end{align}

We now discuss the values of $n_{\pm}$ and $p_1$ for which regular solutions are actually possible.  First of all,  as discussed in section \ref{sec:global},  we need to meet the necessary condition that $q_1\,q_2<0$.  From the definition \eqref{systemflux} of the integers $p_{1,2}$ that characterize the magnetic fluxes,  we notice that
\begin{align}
p_1\,p_2\, = \, \frac{(n_-\,n_+)^2\,(y_2^2-y_3^2)^2}{P(y_2)\,P(y_3)}\left(\frac{\Delta z}{2\pi}\right)^2\,q_1\,q_2\,,
\end{align}
from which it is clear that the product $p_1\,p_2$ has the same sign as the product $q_1\,q_2$ and 
hence in order that a regular solution exists we impose
\begin{align}
p_1 p_2<0\,.
\end{align}
Recalling \eqref{constraintpsen},
we must have
\begin{align}\label{conditionp1}
p_1\, <\, 0 \, , \qquad \text{or} \qquad p_1\, >\, n_-+n_+\,.
\end{align}
While in principle this is only a necessary condition for regularity of our solution,  we find that it guarantees the existence of four real roots for all $n_{\pm}\in \mathbb{N}$,  with $y_{2,3}>0$.  Note in particular that \eqref{conditionp1} guarantees that $\mathtt{s}\in \mathbb{R}$, where 
$\mathtt{s}$ was introduced in \eqref{sdef}.  Moreover,  we have assumed in our analysis that $y_2<y_3$,  which is realized when 
\begin{align}
n_->n_+\,,
\end{align}  
which we shall henceforth assume.  

We also note that both $y_2$ and $y_3$ are symmetric under the exchange $p_1\leftrightarrow p_2=n_-+n_+-p_1$,  so the behaviour on the two branches given in \eqref{conditionp1} is symmetric.  
We find for the allowed values of $p_1$ at fixed $n_+$ that $y_2$ is a monotonically decreasing function of $n_-$,  while $y_3$ is monotonically increasing,  with
\begin{align}
\lim_{n_-\to+\infty}y_2\, = \, 0\,, \qquad
\lim_{n_-\to+\infty}y_3\, = \, -\frac{2\,p_1}{n_+-p_1}\,.
\end{align}
At the two finite extrema $p_1=0$ and $p_1=n_-+n_+$ of the intervals \eqref{conditionp1} we find that $y_{2,3}$ and $q_{1,2}$ all vanish identically,  corresponding to the degenerate case with a double root discussed in section \ref{sec:global}. 

Finally,  we note that in the special case $n_-=n_+=n$ the two roots $y_{2,3}$ become degenerate,  which implies that the space in the $y$ and $z$ direction is non-compact. This can also be seen from the fact that $\Delta z$ diverges in this case --  see equation  \eqref{Deltaz}. In particular,  this implies that there is no limit in which one can obtain a smooth $S^2$ horizon,  which would correspond to $n_-=n_+=1$.

\section{Uplift to M-theory and the central charge}
\label{sec:uplift}

In this section we uplift the $AdS_5\times \Sigma$ gauged supergravity solutions 
of section \ref{sec:soln} to $D=11$ supergravity. We then quantize the 
four-form flux and compute the central charge of the solutions.

\subsection{Uplift to M-theory}

All (supersymmetric) solutions of $D=7$, $U(1)^2$ gauged supergravity can be uplifted to (supersymmetric) solutions of $D=11$ supergravity.  Following \cite{Cvetic:1999xp},  the metric in $D=11$ can be written as
\begin{align}\label{upliftmetric}
L^{-2}\,\diff s^2_{11} &\,  = \, \Omega^{1/3}\diff s^2_7 + \Omega^{-2/3}\Big[X_0^{-1} \diff \mu_0^2 + \sum_{i=1}^2 X_i^{-1}\left(\diff \mu_i^2 + 
\mu_i^2(\diff\phi_i + A_i )^2\right)\Big]\, .
\end{align}
Here 
$\diff s^2_7$ denotes the $D=7$ gauged supergravity metric, it is convenient to introduce $X_0 \equiv (X_1X_2)^{-2}$, and we have defined
the warp factor function
\begin{align}
\Omega  \, \equiv \, \sum_{a=0}^2  X_a\,\mu_a^2\, .
\end{align}
The coordinates $\mu_0,\mu_1,\mu_2$ satisfy the constraint $\sum_{a=0}^2 \mu_a^2=1$, and 
$\phi_1$, $\phi_2$ both have period $2\pi$. These parametrize $S^4\subset \R\times \R^2\times \R^2$, 
where $\mu_0\in\R$, and $(\mu_i,\phi_i)$ form polar coordinates on the two copies of $\R^2$, $i=1,2$. 
The Abelian gauge fields $A_1$, $A_2$ then twist these $\R^2\cong \C$ directions over the $D=7$ spacetime. 
The Hodge dual of the four-form flux  can be expressed as
\begin{align}\label{upliftflux}
L^{-6}\,\ast_{11}G_4\,\,=&\,\,\,2\,\sum_{a=0}^2\,\left(X^2_{a}\,\mu^2_{a}-\Omega\,X_a\right)\,\text{vol}_7+\Omega\,X_0\,\text{vol}_7+\frac{1}{2}\sum_{a=0}^2 X_a^{-1}\,\ast_7 \diff X_a \wedge \diff(\mu^2_a)\nonumber\\
&+\frac{1}{2}\sum_{i=1}^2X_i^{-2}\,\diff(\mu^2_i)\wedge (\diff \phi_i+A_i) \wedge \ast_7 F_i\,,
\end{align}
where $\ast_7$ and $\text{vol}_7$ are the Hodge dual and volume form of the seven-dimensional metric $\diff s^2_7$,  while $\ast_{11}$ is the Hodge dual with respect to the full eleven-dimensional metric $\diff s^2_{11}$.
The $AdS_7$ vacuum solution uplifts to $AdS_7\times S^4$ and is dual to the $d=6$, $\mathcal{N}=(0,2)$ SCFT. The
fact that $\Delta\phi_i=2\pi$ allows us to identify 
\begin{align}\label{defrgf}
A^R\equiv A_1+A_2\,,
\end{align}
as the gauge field associated with the R-symmetry of this SCFT, as mentioned earlier. 

Uplifting the spindle solution given in \eqref{7dmetric} we find that 
the $D=11$ metric takes the form
\begin{align}\label{D11metric}
\diff s^2_{11} &\,  = \,  L^2\mathrm{e}^{2\lambda} \left[\diff s^2_{AdS_5}+\diff s^2_{M_6} \right] \, ,
\end{align}
where
\begin{align}\label{lambda}
\ex^{2\lambda}\, = \, \left(y\,\pp(y)\right)^{1/5}\,\Omega^{1/3}\,,
\end{align}
and the metric on the internal space $M_6$ is
\begin{align}\label{M6metric}
\diff s^2_{M_6}\, = \, \diff s^2_{\Sigma}+\frac{1}{(y\pp(y))^{1/5}\,\Omega}\Big[X_0^{-1} \diff \mu_0^2 + \sum_{i=1}^2 X_i^{-1}\left(\diff \mu_i^2 + 
\mu_i^2(\diff\phi_i + A_i )^2\right)\Big]\,.
\end{align} 
 Recall that the gauge field fluxes through the spindle are
\begin{align}
P_i \, = \, \frac{1}{2\pi}\int_\Sigma \diff A_i \, = \, \frac{p_i}{n_-n_+}\, .
\end{align}
Provided $p_i\in \Z$ are integers, each of which is relatively prime to both $n_-$ and $n_+$, then 
 \eqref{M6metric} is (with the exception of certain orbifold singularities we describe below) a smooth metric on the total space of an $S^4$ bundle 
over $\Sigma$:
\begin{align}\label{fibration}
S^4\ \hookrightarrow \ M_6 \ \rightarrow \ \Sigma\, . 
\end{align}
More precisely, notice that $\mu_0\in [-1,1]$, with $\mu_0=\pm 1$ corresponding to the north 
and south poles of the $S^4$ internal space. At these points the copies of $S^3$ parametrized by 
$(\mu_i,\phi_i)$, with $\mu_1^2+\mu_2^2=1-\mu_0^2$, collapse to a point. For fixed $\mu_0\in (-1,1)$ 
the resulting $S^3$ fibrations over $\Sigma$ are completely smooth manifolds. However, 
notice that the total space $M_6$ has orbifold singularities at the poles $\mu_0=\pm 1$: each 
of these is a copy of the spindle $\Sigma$, and the latter has orbifold singularities. 
The $D=11$ solution is thus mildly singular.

The $D=11$ four-form flux $G_4$ is closed. Since as a topological space 
$\Sigma$ is homeomorphic to a two-sphere, the integral of $G_4$ through any $S^4$ 
fibre of \eqref{fibration}, at any fixed point on the spindle $\Sigma$, will be independent of 
the latter choice of point. We interpret this flux number
\begin{align}
N \, = \, \frac{1}{(2\pi \ell_p)^3}\int_{S^4}G_4\, ,
\end{align}
as the number of M5-branes wrapped on $\Sigma$. 
Indeed, recalling that 
\begin{align}\label{TTwist}
P_1 + P_2 \, = \, \chi(\Sigma) \, = \ \frac{n_-+n_+}{n_-n_+}\, ,
\end{align}
where $\chi(\Sigma)$ is the Euler number of the spindle, this solution 
has the following natural interpretation. One begins with the 
supersymmetric solution
$\R^{1,3}\times \R \times Y_6$ of $D=11$ supergravity, where 
$Y_6$ is a local Calabi-Yau three-fold of the form
\begin{align}
Y_6\, = \, \mathcal{N}_1 \oplus \mathcal{N}_2\ \rightarrow \ \Sigma\, .
\end{align}
Here $\mathcal{N}_i$ are the two complex line bundles on which $-A_i$ are Hermitian connections, 
so that $\mathcal{N}_i=\mathcal{O}(-p_i)$. By virtue of \eqref{TTwist}, 
the total space of $Y_6$
has vanishing first Chern class and hence is indeed a Calabi-Yau three-fold. One then wraps $N$ M5-branes 
over the zero section $\Sigma$ of $Y_6$, which is a holomorphic curve,
and at the origin in the copy of 
$\R$. At low energies the effective theory on the M5-branes 
will be a $d=4$, $\mathcal{N}=1$ supersymmetric field
theory on the Minkowski space $\R^{1,3}$. 

In this construction, supersymmetry on the wrapped M5-branes in the UV
is being preserved by a topological twist, although 
as we already noted in section \ref{sec:global} the metric on $\Sigma$ 
in the IR does not (and indeed \emph{cannot}) have constant curvature.
This is in constrast to the case studied in \cite{Anderson:2011cz}, 
where $\Sigma=\Sigma_{\mathtt{g}}$ is a \emph{smooth} 
Riemann surface of genus $\mathtt{g}$, without orbifold singularities. 
In the latter case the authors show that the metric on $\Sigma$ in the IR 
necessarily flows to a constant curvature metric. The solutions 
in this paper are counterexamples to this result in the case of orbifolds. 

Assuming the compactified $d=4$, $\mathcal{N}=1$ low-energy theory on the M5-branes flows to a SCFT in the IR, 
our $D=11$ solution \eqref{D11metric} is then naturally 
interpreted as the near horizon limit of this 
system of $N$ M5-branes wrapped on $\Sigma$.
This IR solution, which does not exhibit a (local) topological twist with Killing spinors that are constant on $\Sigma$, as noted earlier,  is holographically dual to the associated $d=4$, $\mathcal{N}=1$ 
SCFT. We shall analyse this dual field theory directly 
in section \ref{sec:anom}, and give further evidence for this claim. 

\subsection{Central charge}

Let us now turn to the supergravity computation of the central charge, $a$, of the $d=4$, $\mathcal{N}=1$ 
SCFT that is dual to the configuration of M5-branes wrapped on a spindle that we have described in this paper.  
We can follow \cite{Gauntlett:2006ai},  where given the form \eqref{D11metric} of the $D=11$ metric it was shown that the central charge can be expressed as
\begin{align}\label{centralchargefrom11d}
a &\,  = \,  \frac{1}{2^7\pi^6} \left(\frac{L}{\ell_p}\right)^9\int_{M_6} \diff^6x\,\sqrt{g_{M_6}}\, \mathrm{e}^{9\lambda}\,, 
\end{align}
where $\ell_p$ is the eleven-dimensional Planck length.
This expression can be computed directly using the values of $\lambda$ and the metric on $M_6$ given in \eqref{lambda} and \eqref{M6metric}.  To this end,  one can solve the constraint $\sum_{a=0}^2\mu_a^2=1$ in terms of two angles $\eta$ and $\theta$,  with 
\begin{align}
\mu_0\, = \, \cos\theta\,, \quad
\mu_1\, = \, \sin\theta\,\cos\eta\,,\quad
\mu_2\, = \, \sin\theta\,\sin\eta\,,
\end{align}
where for $\theta\in [0,\pi]$, $\eta \in [0,\tfrac{\pi}{2}]$ and $\phi_i\in [0,2\pi)$,  the four angles parametrize an $S^4$.  In terms of these,  one finds
\begin{align}
\sqrt{g_{M_6}}\,\ex^{9\lambda}\, = \, \frac{y}{4}\,\sin 2\eta\,\sin^3 \theta\,,
\end{align}
from which it follows that 
\begin{align}\label{atry1}
a\, = \, \frac{L^9}{2^7\pi^6\ell_p^9}\,\text{vol}(S^4)\,\int\frac{y}{2}\,\diff y\, \diff z\, = \, \frac{L^9}{2^7\pi^6\ell_p^9}\,\text{vol}(S^4)\,\frac{y_3^2-y_2^2}{4}\,\Delta z\,,
\end{align}
where 
\begin{align}
\text{vol}(S^4)\, = \, \frac{8\pi^2}{3}\,.
\end{align}
To complete the computation we need an expression for $L$ which comes from quantization 
of the four-form flux $G_4$. In particular, as described in the previous subsection,
we can interpret  
\begin{align}
N \, = \, \frac{1}{(2\pi \ell_p)^3}\int_{S^4}G_4\, ,
\end{align}
as the number of M5-branes wrapped on $\Sigma$.  This leads to
\begin{align}
L\, = \, (\pi N)^{1/3}\,\ell_p\,.
\end{align}
Using this, as well as \eqref{y_{2,3}} and \eqref{Deltaz} in \eqref{atry1},
we obtain the following result for the central charge of the dual SCFT
\begin{align}\label{sugracentralcharge}
a \, &  = \,\frac{3\,p_1^2\,p_2^2\,(\mathtt{s}+p_1+p_2)}{8\,n_-\,n_+\,(n_--p_1)\,(p_2-n_-)\,[\mathtt{s}+2\,(p_1+p_2)]^2}\,N^3\, ,
\end{align}
where $p_2=n_-+n_+-p_1$.  In the next section  we shall see that this is exactly reproduced by a field theory computation,  using the anomaly polynomial of the theory on the M5-branes. 

\section{Field theory}\label{sec:anom}

\subsection{M5-brane anomaly polynomial}\label{sec:M5poly}

To leading order in the large $N$ limit, 
the anomaly polynomial for $N$ M5-branes is given by the eight-from
\begin{align}\label{A6}
\mathcal{A}_{\mathrm{6d}} \, =\, \frac{1}{24}p_2(R) N^3 +{O}(N)\, ,
\end{align}
(see, for example, \cite{Hosseini:2020vgl}).
Here $R$ denotes the $SO(5)_R$ symmetry of the worldvolume theory, which 
geometrically rotates the normal directions to the M5-branes in spacetime, and $p_2(R)$ is the second 
Pontryagin class. This $SO(5)_R$ then also rotates the $S^4$ internal space 
in the uplift from $D=7$ gauged supergravity to $D=11$. The ${O}(N)$ corrections in \eqref{A6} involve Pontryagin classes 
also of the tangent bundle of the worldvolume. It is straightforward to keep these sub-leading corrections in the following analysis, but we do not do so both because the results are rather lengthy and also because they are not
needed in comparing with the leading order supergravity result for the $a$ central charge. 

Recall that the two Abelian gauge fields $A_i$, $i=1,2$, of $D=7$ gauged supergravity are connections 
for the Cartan subgroup $U(1)\times U(1)\subset SO(5)_R$. The normal bundle $\mathcal{N}$ to the M5-branes 
wrapped on $\R^{1,3}\times \Sigma$ is 
\begin{align}
\mathcal{N}\, = \, \R\oplus \mathcal{N}_1\oplus \mathcal{N}_2\, ,
\end{align}
where $\mathcal{N}_i$ are complex line bundles on which $-A_i$ are the connections. 
The Pontryagin class in \eqref{A6} may then be written in terms of first Chern classes, so that
\begin{align}\label{A6again}
\mathcal{A}_{\mathrm{6d}} \, =\, \frac{1}{24}c_1(\mathcal{N}_1)^2c_1(\mathcal{N}_2)^2 N^3\, , 
\end{align}
where we will suppress the ${O}(N)$ corrections in what follows, writing only the leading order result.

We want to compactify the worldvolume theory of the M5-branes on $\Sigma=\mathbb{WCP}^1_{[n_-,n_+]}$, where the 
fluxes that twist the normal directions satisfy 
\begin{align}
\int_\Sigma c_1(\mathcal{N}_i) \, = \, -\frac{1}{2\pi}\int_\Sigma \diff A_i\, \ = \, -P_i\, .
\end{align}
As commented earlier, the fact that $P_1+P_2=\chi(\Sigma)$ is equivalent to the total space of 
$\mathcal{N}_1\oplus\mathcal{N}_2\rightarrow \Sigma$ being a Calabi-Yau three-fold. 
In compactifying the $d=6$ theory to $d=4$ on the spindle $\Sigma$, we need to take into 
account the $U(1)_{\mathcal{J}}$ global symmetry in $d=4$ that arises from the 
isometry of $\Sigma$. As explained in  \cite{Ferrero:2020laf} (see also \cite{Hosseini:2020vgl}, \cite{Bah:2019rgq}), 
we then wish to compute the anomaly polynomial \eqref{A6again}, where the 
eight-manifold $Z_8$ on which it is defined is the total space of a $\Sigma$ 
fibration over $Z_6$, so
\begin{align}\label{Zfibre}
\Sigma \ \hookrightarrow \ Z_8 \ \rightarrow \ Z_6\, .
\end{align}
As in \cite{Ferrero:2020laf}, in order to define this twisting 
it is important to ensure that the $D=7$ Killing spinor is invariant 
under the $U(1)_{\mathcal{J}}$ symmetry generated by 
$\partial_\varphi$, where we have defined $\varphi\equiv \frac{2\pi}{\Delta z}z$, with 
$\Delta\varphi=2\pi$. 
The explicit Killing spinor for a Wick rotation of these solutions 
was constructed in \cite{Liu:1999ai}, and in our notation 
this is independent of $z$ provided we make the 
gauge transformation
\begin{align}
A_i \ \to \  \tilde{A}_i\, = \, A_i -\frac{1}{4}\,\diff z\,.
\end{align}
We then note that the R-symmetry gauge field, $A^R \, \equiv \, A_1 + A_2$, 
satisfies
\begin{align}\label{gaugeconditions}
\left.\tilde{A}^R\right|_{y=y_2}\, = \, -\frac{1}{n_+}\,\diff\varphi\, ,\qquad
\left.\tilde{A}^R\right|_{y=y_3}\, = \, +\frac{1}{n_-}\,\diff\varphi\,, 
\end{align}
at the poles of the spindle $\Sigma$. 
In this gauge\footnote{Notice that we are working in a specifc gauge for the ``flavour" symmetry gauge field, $A_1-A_2$, that is given by the original supergravity solution. This is analogous to the analysis of D3-branes wrapped on a spindle in \cite{Boido:2021szx}, whereas in the same setting other gauge choices were analysed in
\cite{Hosseini:2021fge}, leading to the same final result for the central charge.} we then replace 
$\diff\varphi\rightarrow\diff\varphi+\mathcal{A}_{\mathcal{J}}$, and correspondingly introduce connection 
one-forms on $Z_8$ 
\begin{align}
\mathscr{A}_i\, = \, \left(\frac{q_i}{h_i(y)}-\frac{1}{4}\right)\frac{\Delta z}{2\pi}(\diff \varphi +\mathcal{A}_{\mathcal{J}})\, \equiv \, \rho_i(y)\,(\diff \varphi +\mathcal{A}_{\mathcal{J}})\,,
\end{align}
for $i=1,2$, with curvature
\begin{align}
\mathscr{F}_i\, = \, \diff \mathscr{A}_i=\rho_i'(y)\,\diff y\wedge(\diff \varphi +\mathcal{A}_{\mathcal{J}})+\rho_i(y)\,\mathcal{F}_{\mathcal{J}}\,,
\end{align}
where $\mathcal{F}_{\mathcal{J}}\equiv \diff \mathcal{A}_{\mathcal{J}}$. These have the property that $\mathscr{A}_i$  
restrict to the supergravity gauge fields $\tilde{A}_i$ on each $\Sigma$ fibre of \eqref{Zfibre}. 
We then write the first Chern classes  $c_1(\mathcal{L}_i)=[\mathscr{F}_i/2\pi]\in H^2(Z_8,\R)$,  $c_1(\mathcal{J})=[\mathcal{F}_J/2\pi]\in H^2(Z_6,\Z)$, and 
write
\begin{align}
c_1(\mathcal{N}_i)\, = \,  \Delta_i\,c_1(R_{\mathrm{4d}})-c_1(\mathcal{L}_i)\,.
\end{align}
Here $R_{\mathrm{4d}}$ is the pull-back of a $U(1)_R$ symmetry bundle over $Z_6$, and the trial R-charges 
$\Delta_i$ satisfy 
\begin{align}\label{Deltacon}
\Delta_1+\Delta_2\, = \, 2\, , 
\end{align}
which is the constraint that the preserved spinor has R-charge 1.

The $d=4$ anomaly polynomial is then obtained by integrating $\mathcal{A}_{\mathrm{6d}}$ in \eqref{A6again} 
over $\Sigma$,
\begin{align}
\mathcal{A}_{\mathrm{4d}}\, = \, \int_{\Sigma}\mathcal{A}_{\mathrm{6d}}\,,
\end{align}
which gives 
\begin{align}\label{A4}
\mathcal{A}_{\mathrm{4d}} \, = \, \Big[&
-\Delta_1\Delta_2\,(\Delta_1 I_1+\Delta_2 I_2)\,c_1(R_{\mathrm{4d}})^3-(\Delta_1 I_6+\Delta_2 I_7)\,c_1(R_{\mathrm{4d}})\,c_1(\mathcal{J})^2\nn
& +(\Delta_1^2 I_3  +2\Delta_1\Delta_2 I_4+\Delta_2^2 I_5)\,c_1(R_{\mathrm{4d}})^2\,c_1(\mathcal{J})+I_8\, c_1(\mathcal{J})^3\Big]\frac{N^3}{24}\,.
\end{align}
Here $I_\alpha$, $\alpha=1,\ldots,8$, are certain integrals of $\rho_i(y)$ and their deriatives $\rho_i'(y)$, which 
are reported in the appendix. 

\subsection{$a$-maximization}\label{sec:amax}

Having obtained the $d=4$ anomaly polynomial \eqref{A4}, it is now straightforward to extract the trial 
$a$ central charge. Specifically, the coefficient of $\frac{1}{3!}c_1(L_i)c_1(L_j)c_1(L_k)$ in 
$\mathcal{A}_{\mathrm{4d}}$ computes the trace $\mathrm{Tr}\, \gamma^5 Q_iQ_jQ_k$, 
where the global symmetry $Q_i$ is associated to the complex line bundle $L_i$ over 
$Z_6$. In the large $N$ limit the trial $a$ central charge is then 
\begin{align}
a_{\mathrm{trial}} \ = \ \frac{9}{32}\mathrm{Tr}\, \gamma^5 R^3_{\mathrm{trial}}\, ,
\end{align}
where we allow for a mixing with the $U(1)_{\mathcal{J}}$ global symmetry by taking
\begin{align}
R_{\mathrm{trial}}\, = \, R_{\mathrm{4d}} + \varepsilon\, \mathcal{J}\, .
\end{align}
This leads to the trial $a$ function
\begin{align}\label{atrial}
a_{\mathrm{trial}}\, = \, \frac{9}{32} \frac{N^3}{24}\,3!\,\Big[ &-\Delta_1\, \Delta_2\,(\Delta_1\, I_1+\Delta_2\, I_2)+(\Delta_1^2\, I_3+2\,\Delta_1\, \Delta_2\, I_4+\Delta_2^2 \,I_5)\,\varepsilon \nn
& -(\Delta_1 I_6+\Delta_2 I_7)\,\varepsilon^2+I_8\,\varepsilon^3\Big]\,. 
\end{align}
Here the choice of R-symmetry is parametrized by $\varepsilon$ and $\Delta_1,\Delta_2$, where the latter are 
subject to the constraint \eqref{Deltacon}. The exact superconformal 
R-symmetry locally maximizes \eqref{atrial} \cite{Intriligator:2003jj}, and we find the 
extremal values
\begin{align}
\Delta_1^*\, = \, \Delta_2^*\, = \, 1\,,
\end{align}
together with 
\begin{align}\label{extremaleps}
\varepsilon^* 
\, &= \,
\frac{n_-\,n_+\,(2\,n_--p_1-p_2)(\mathtt{s}+p_1+p_2)}{(n_--p_1)(p_2-n_-)[\mathtt{s}+2(p_1+p_2)]}\,,
\end{align}
where recall from \eqref{sdef} that 
\begin{align}
\mathtt{s} \, = \, \sqrt{7\,(p_1^2+p_2^2)+2\,p_1\,p_2-6\,(n_-^2+n_+^2)}\, .
\end{align}

Notice that the superconformal R-symmetry is then
\begin{align}\label{rsymscft}
R^* \, = \, R_{\mathrm{4d}} + \varepsilon^*\, \mathcal{J}\, ,
\end{align}
which thus mixes non-trivially with the $U(1)_\mathcal{J}$ isometry of the spindle. 
Interestingly, we find that 
\begin{align}\label{matchR}
\varepsilon^*\, = \, \frac{4}{3}\left(\frac{2\pi}{\Delta z}\right)\, ,
\end{align}
with $\Delta z$ computed for the gravity solution in \eqref{Deltaz}. We expect equation \eqref{matchR}
 to be crucial for matching the superconformal R-symmetry, computed in field theory 
in this section, with the superconformal R-symmetry as realized in the supergravity 
dual. The latter should be constructed as a Killing vector bilinear in the $D=11$  Killing spinor, 
as in \cite{Ferrero:2020twa}, \cite{Ferrero:2020laf}, and indeed an analogous 
equation to \eqref{matchR} was precisely used in the latter reference to get such an agreement. 
However, we leave this detail for future work.

The $a$ central charge of the $d=4$, $\mathcal{N}=1$ SCFT is then the extremal value 
\begin{align}\label{ftheorycentchge}
a \, &  =\, a_{\mathrm{trial}}(\varepsilon^*,\Delta_1^*,\Delta_2^*) \nn [6pt]
&=\frac{3\,p_1^2\,p_2^2\,(\mathtt{s}+p_1+p_2)}{8\,n_-\,n_+\,(n_--p_1)\,(p_2-n_-)\,[\mathtt{s}+2\,(p_1+p_2)]^2}\,N^3\,,
\end{align}
where recall that $p_2=n_-+n_+-p_1$.
This agrees perfectly with the supergravity result given in \eqref{sugracentralcharge}.

Note that for certain values of $n_\pm,p_1$ the $a$ central charge can be a rational number.
For such values
the superconformal R-symmetry \eqref{rsymscft} is a compact $U(1)$ symmetry, while generically it is non-compact.
For example, $a$ is rational for
$n_-=2$, $n_+=1$, $p_1=(-1,-8,-34,...)$ and
$n_-=3$, $n_+=2$, $p_1=(-1,-4,-11,...)$ as well as replacing $p_1\to n_-+n_+-p_1$. 
In fact, a subset of cases are given by the analytic formula
\begin{align}
p_1=\,&\,\frac{n_-+n_+}{2}-\frac{3\,n_--n_+}{4}\left[\beta_+^{k}+\beta_-^{k}\right]
-\frac{5\,n_--n_+}{4\sqrt{3}}\left[\beta_+^{k}-\beta_-^{k}\right]\,,
\end{align}
where $k=0,1,2,\dots$ and $\beta_\pm=2\pm\sqrt{3}$.

It is also interesting to note that if we formally set $n_+=n_-=1$ in \eqref{ftheorycentchge} then we get
$a=\frac{1-9z^2+(1+3z^2)^{3/2}}{48 z^2}N^3$
where $z=1-p_1$, which exactly agrees with the field theory result given in eq. (2.22) of \cite{Bah:2012dg} for
M5-branes wrapping a Riemann surface of genus $g=0$ with a standard topological twist. We also observe that 
formally setting $n_+=n_-=1$ in \eqref{extremaleps} gives $\varepsilon^*=0$ associated with no mixing with the 
(non-abelian) isometries of the round two-sphere, as in \cite{Bah:2012dg}.

\section{Discussion}
\label{discuss:sec}

Starting with the seminal work of \cite{Maldacena:2000mw}, there is now
a plethora of supersymmetric $AdS$ solutions that are holographically dual to branes wrapping cycles. In these constructions
 supersymmetry is typically realized via a topological twist,
with Killing spinors that are constant on the cycle.
In    \cite{Ferrero:2020laf}  it was demonstrated  that there exists a novel realization of supersymmetry when D3-branes wrap a spindle, with the Killing spinors becoming sections of a
 non-trivial bundle. This discovery, together with an analogous construction for M2-branes wrapped on a spindle \cite{Ferrero:2020twa}, indicate that there exists a rich new landscape of supergravity constructions representing branes wrapped on spindles, along with their associated field theory duals.

In this paper we continued to explore this landscape, concentrating on a new class of supersymmetric solutions describing M5-branes wrapping spindles. The $AdS_5\times \Sigma$ solutions are constructed in $D=7$ gauged supergravity and then uplifted to $D=11$.
While there are similarities with the analogous solutions
describing D3-branes and M2-branes wrapped on a spindle \cite{Ferrero:2020laf,Ferrero:2020twa},
the details of how supersymmetry is realized are novel. Once again, it is not via the standard topological twist, 
since the R-symmetry gauge field does not cancel the spin connection, and as a consequence the Killing spinors are not constant on the spindle. However, globally there is a kind of topological twist, 
in the sense that the R-symmetry flux is equal to the Euler character of the spindle.

The results of this paper, together with \cite{Ferrero:2020laf,Ferrero:2020twa}, open up several research directions.  
An immediate question is to understand the reason why supersymmetry is being realized in a different fashion for
the D3-branes and M2-branes wrapping a spindle than it is for M5-branes. It may be possible to realize
supersymmetry in other ways too. It would also be of interest to demonstrate explicitly how our new solutions fit in the general classification of $AdS_5\times M_6$ solutions of $D=11$ supergravity given in \cite{Gauntlett:2004zh}.
Investigating constructions that preserve ${\cal N}=2$
supersymmetry\footnote{We pointed out in section \ref{sec:global} that $q_1q_2=0$ should give rise to 
${\cal N}=2$
supersymmetric solutions but not with a compact spindle. This is also directly related to the fact that the central charge
\eqref{sugracentralcharge} vanishes for $p_1p_2=0$.
We also note that as we were finalizing this paper, the paper \cite{Bah:2021mzw} appeared on the arXiv, where 
work on ${\cal N}=2$ solutions is reported, using the same local family of solutions.} would be worthwhile and connecting with the work of \cite{Gaiotto:2009gz}.
In another direction, we expect generalizations involving branes wrapping various higher-dimensional cycles equipped with orbifold metrics.

In further explorations of the landscape of supergravity solutions wrapping orbifolds, it will be illuminating to
utilize techniques that do not require finding the solutions in explicit form, such as those developed in
 \cite{Couzens:2018wnk,Gauntlett:2019roi,Gauntlett:2019pqg}, following the paradigm of \cite{Martelli:2005tp,Martelli:2006yb}. In particular, the approach of  \cite{Couzens:2018wnk,Gauntlett:2019roi,Gauntlett:2019pqg} does not require making any
 \emph{a priori} assumption on the metrics. In fact, in \cite{Gauntlett:2019pqg} key properties  
 of the solutions discussed in  \cite{Ferrero:2020laf,Ferrero:2020twa}, 
 such as the central charge and the entropy, respectively,
were obtained using the geometric extremization principle of \cite{Couzens:2018wnk}.

\section*{Acknowledgments}

\noindent 
D.M. thanks F.  Faedo for useful discussions. 
This work was supported in part by STFC grants ST/T000791/1 and 
ST/T000864/1. 
JPG is supported as a Visiting Fellow at the Perimeter Institute. 

\appendix
\section{Integral coefficients}
In this appendix we give the explicit values of the integrals $I_\alpha$ introduced in section \ref{sec:anom}.  These arise in the computation of the four-dimensional anomaly polynomial $\mathcal{A}_{\mathrm{4d}}$, 
and explicitly,  they are given by
\begin{align}
I_1&\, = \, \frac{1}{\pi}\int_{\Sigma}\frac{\diff}{\diff y}\rho_2(y)\,\diff y \,\diff \varphi\, = 2\,\left(\rho_2(y_3)-\rho_2(y_2)\right)\,, \nn
I_2&\, = \, \frac{1}{\pi}\int_{\Sigma}\frac{\diff}{\diff y}\rho_1(y)\,\diff y \,\diff \varphi\, = \, 2\,\left(\rho_1(y_3)-\rho_1(y_2)\right)\,, \nn
I_3&\, = \, \frac{1}{2\pi}\int_{\Sigma}\frac{\diff}{\diff y}\rho_2(y)^2\,\diff y \,\diff \varphi\, = \, \rho_2(y_3)^2-\rho_2(y_2)^2\,,\nn
I_4&\, = \, \frac{1}{\pi}\int_{\Sigma}\frac{\diff}{\diff y}\rho_1(y)\,\rho_2(y)\,\diff y \,\diff \varphi\, = \, 2\,\left(\rho_1(y_3)\,\rho_2(y_3)-\rho_1(y_2)\,\rho_2(y_2)\right)\,,\nn
I_5&\, = \, \frac{1}{2\pi}\int_{\Sigma}\frac{\diff}{\diff y}\rho_1(y)^2\,\diff y \,\diff \varphi\, = \, \rho_1(y_3)^2-\rho_1(y_2)^2\,,\nn
I_6&\, = \, \frac{1}{\pi}\int_{\Sigma}\frac{\diff}{\diff y}\left[\rho_2(y)^2\,\rho_1(y)\right]\,\diff y\,\diff \varphi\,= \,2\,\left(\rho_2(y_3)^2\,\rho_1(y_3)-\rho_2(y_2)^2\,\rho_1(y_2)\right)\,,\nn
I_7&\, = \, \frac{1}{\pi}\int_{\Sigma}\frac{\diff}{\diff y}\left[\rho_1(y)^2\,\rho_2(y)\right]\,\diff y\,\diff \varphi\nonumber\, = \, 2\,\left(\rho_1(y_3)^2\,\rho_2(y_3)-\rho_1(y_2)^2\,\rho_2(y_2)\right)\,,\nn 
I_8&\, = \, \frac{1}{2\pi}\int_{\Sigma}\frac{\diff}{\diff y}\left[\rho_1(y)^2\,\rho_2(y)^2\right]\,\diff y\,\diff \varphi\, = \, \rho_1(y_3)^2\,\rho_2(y_3)^2-\rho_1(y_2)^2\,\rho_2(y_2)^2\,,
\end{align}
where
\begin{align}
\rho_1(y_2)&=-\frac{(n_-+n_+-\mathtt{s})\,(2\,n_-^2+2\,n_-\,n_+-3\,n_-\,p_1-n_+\,p_1+2\,p_1^2+n_-\,\mathtt{s})}{12\,n_-\,n_+\,(n_--n_+)\,(n_--p_1)}\,,\nonumber\\
\rho_1(y_3)&=-\left. \rho_1(y_2)\right|_{n_-\leftrightarrow n_+}\,, \nn
\rho_2(y_2)&=\left. \rho_1(y_2)\right|_{p_1\leftrightarrow p_2}\,, \nn
\rho_2(y_3)&=-\left. \rho_2(y_2)\right|_{n_-\leftrightarrow n_+}\,.
\end{align}

%\bibliographystyle{utphys}
%\bibliography{helical}{}

\providecommand{\href}[2]{#2}\begingroup\raggedright\endgroup

\end{document}